# Robust semi-automatic vessel tracing in the human retinal image by an instance segmentation neural network

**Siyi Chen[1], Amir H. Kashani[2], Ji Yi[1,2, *]**

[1.] *Department of Biomedical Engineering, Johns Hopkins University, Baltimore MD, 21231*

[2.] *Department of Ophthalmology, Johns Hopkins University, School of Medicine, Baltimore MD, 21231*

*Corresponding author: jiyi@jhu.edu

**Abstract**: The morphology and hierarchy of the vascular systems are essential for perfusion in supporting metabolism. In human retina, one of the most energy-demanding organs, retinal circulation nourishes the entire inner retina by an intricate vasculature emerging and remerging at the optic nerve head (ONH). Thus, tracing the vascular branching from ONH through the vascular tree can illustrate vascular hierarchy and allow detailed morphological quantification, and yet remains a challenging task. Here, we presented a novel approach for a robust semi-automatic vessel tracing algorithm on human fundus images by an instance segmentation neural network (InSegNN). Distinct from semantic segmentation, InSegNN separates and labels different vascular trees individually and therefore enable tracing each tree throughout its branching. We have built-in three strategies to improve robustness and accuracy with temporal learning, spatial multi-sampling, and dynamic probability map. We achieved 83% specificity, and 50% improvement in Symmetric Best Dice (SBD) compared to literature, and outperformed baseline U-net. We have demonstrated tracing individual vessel trees from fundus images, and simultaneously retain the vessel hierarchy information. InSegNN paves a way for any subsequent morphological analysis of vascular morphology in relation to retinal diseases.

# Introduction

In the retina, vascular alterations have been associated with a broad range of important pathologies that affect millions worldwide, such as diabetic retinopathy, age-related macular degeneration (AMD), and glaucoma. Morphological and developmental abnormality, including changes of vessel branching, length, caliber, tortuosity at different levels of retinal circulation, can directly impact the retinal perfusion and result from retinal tissue damage. [1, 2] Therefore, quantification of vascular structures is essential in understanding pathogenesis, early detection, and progression in retina diseases.

The vessels in the retinal circulation are "rooted" from optic nerve head (ONH), analogous to a tree, and branch out in the inner retina to construct a vascular perfusion system. In the human retina, there are several pairs of arteries and veins centered at ONH, each of which develops into their individual trees. Being able to follow the vessels from the roots along the branching, and record the hierarchy of vasculature is a key step in characterizing vascular morphology. We call this task by vessel tracing. It is significantly more difficult than vessel segmentation [3-6], and classification [7, 8], both of which do not trace the vascular branching or produce vessel hierarchy. The challenge is further exacerbated when vessels from different trees cross each other in an unpredictable fashion.

There is sparse prior work on vessel tracing, and the state-of-the-art methods underperform. Algorithmic approaches [9-12] have been reported based on skeletonization of the vessels, where each vessel node is classified as a crossing point or bifurcation point according to the vessel angles, and calibers *etc*. However, those methods highly depend on the quality of the skeletonization, and the tailored algorithm has a high rate of failure separating two crossing vessel trees. A different approach [13] uses embedding-based neural network [14] that groups vessel segments into different instances, and differentiate bifurcation points and overlap region. However, they still depend on the skeletonization, and does not produce vascular tracing.

To fill the gap of retina vessel tracing, we regarded vessel tracing as a task of instance segmentation [15], which identifies and outputs different vessel trees individually. This is distinct from conventional semantic segmentation, which only segments out vessels all together from the background. Here, we presented a novel embedding-based neural network (InSegNN) with a discriminative loss function [14], to generate a unique hyper-dimension space where vessel instance clusters are easily separated by clustering algorithms (mean-shift algorithm [16]). Built upon the instance segmentation of vessels, we further developed a robust algorithm to trace individual vascular trees, by incorporating three novel strategies, namely **temporal learning, spatial multi-sampling, and dynamic probability map.**

Using all these approaches, we accomplished a semi-automated and robust vessel tracing algorithm on large-scale retinal fundus images by only manually inputting starting points of vessel trees. For the first time, to our best knowledge, vessel instance segmentation was achieved on fundus images and reached a high accuracy.

# Results

## 2.1 Overview of tracing pipeline

The overall vessel tracing process from a fundus retinal image is illustrated in **Fig. 1a**. To start each trace, a manually annotated "root" vector (blue cross and arrow in **Fig. 1a**) defines the starting point, and sets up the center of the initial "patch". With each patch, we define one instance as a vessel segment that is physically disconnected with others. Note that vessel instances allow branching within the patch. We then used InSegNN to separately label vessel instances. After instance segmentation within the patch, the new set of starting points are enumerated at the patch boundary intersecting only with the designated tree. The "patch" then moves and iterates through all new starting points, until no more new ones found **(Method section 4.3)**. During each iteration, we update the probability map for the designated vessel tree segmentation, and this serves as the template to discard vessels from other trees. The final output separated all trees, and more importantly all the branching sequence since we traced through the vessel structural evolution (**Fig. 1b**).

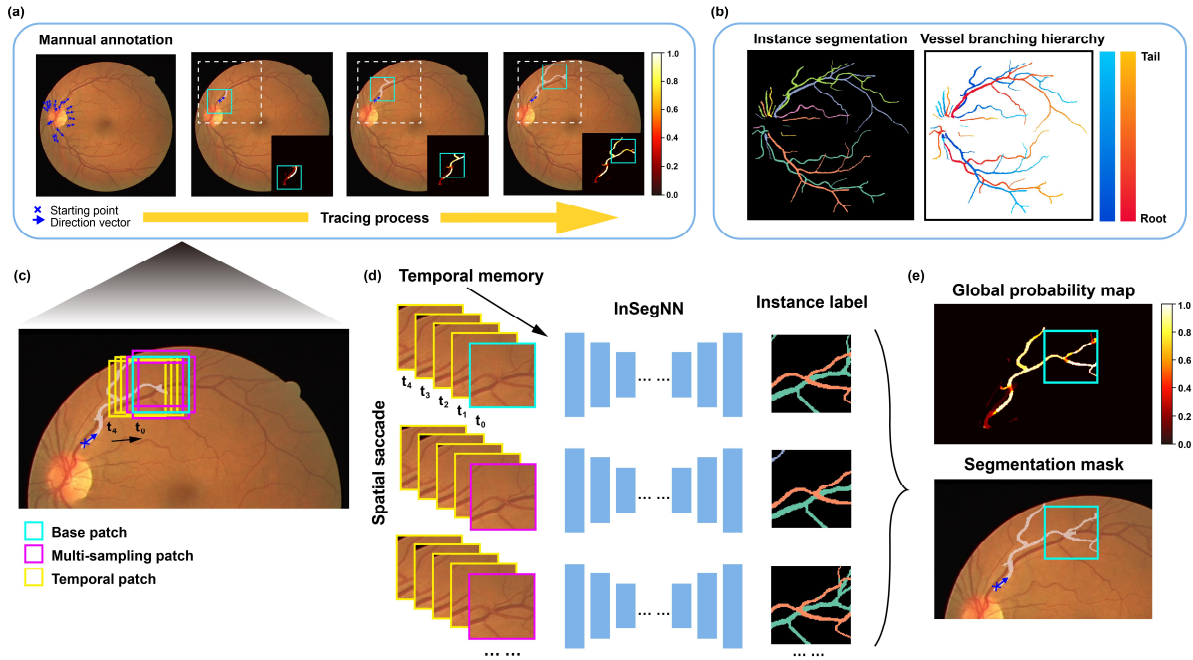

**Figure 1** Overview of the presented vessel tracing method. (a) Flow process of the vessel tracing. Blue arrows are the starting vectors to initiate tracing of each vessel tree from optic nerve head. A moving patch (cyan square) applies instance segmentation and updates a probability map for tracing all branching of the designated tree (zoom-in view). (b) Outcomes of tracing, consisting of vascular instance segmentation and branching hierarchy of the tree structure. (c) The illustration of input for InSegNN. Cyan square is base patch; magenta squares are multi-sampling patches by spatial shifting around the base patch; yellow squares are temporal patches used for temporal learning. Base patch and multi-sampling patch are served as last frames ($t_0$) of temporal sequences. Temporal patches constitute 4 frames ($t_1$ to $t_4$) of sequences, where $t_4$ is the earliest frame along tracing. (d) Five temporal patches ($t_0$-$t_4$) are input of InSegNN. Replacing $t_0$ by multi-sampling patches produces five parallel instance segmentation stacks. (e) A global probability map summarizes the multi-sampling statistics, and produces robust tracing by generating a segmentation mask with a threshold of 0.6.

The major innovation is that InSegNN circumvents the traditional algorithmic approach to analyze vessel crossing. Guided by the discriminative loss function [14] **(Method section 4.1.1)**, the neural network attempts to create a hyper-dimension space where pixels of same instance aggregate while pixels belonging to different instances stay away from each other. This artificial hyper-dimension space, namely embedding space, is an ideal environment for clustering. By using a mean-shift clustering algorithm, pixels in embedding space are labeled according to which instance they belong to, and it is flexible to different instance numbers within a patch.

To augment the data and improve the robustness, we implemented **temporal learning, spatial multi-sampling** in our network. The temporal learning is inspired by a short memory that retains the temporal information [17, 18]. For each new patch, rather than just using this single "base patch" (**Fig. 1c**,**cyan frame**), we packaged it as a movie with a series of retrospective patches (**Fig. 1c, yellow frames**) to help the model learn continuous information of vessels during tracing. The spatial multi-sampling is motivated by microsaccades [19, 20], an involuntary physiological eye movement. It is achieved by slightly moving the base patch multidirectionally (**Fig. 1c magenta frames**) and redoing the model prediction several times in parallel (**Fig. 1d**). The spatial multi-sampling generates the probability map (**Fig. 1e**), which could be binarized for segmentation mask. The probability map is updated by spatial multi-sampling, and during the iteration of new patches. We call it **dynamic probability map** that raise the accuracy of segmentation and significantly improve the robustness.

## 2.2 Instance segmentation neural network

Our InSegNN model outputs instance segmentation labels for all vessels within one patch (**Fig. 2a**). The encoder extracts multiple features of vessels in hierarchic scales while decoder transforms pixels from image space to embedding space, whose dimension is set as 12. The pixels in embedding space form clusters, and can be separated by post-processing. Residual block [21] (**Fig. 2b**) replaces the original block located in all layers in U-Net to lessen over-fitting and improve model performance.

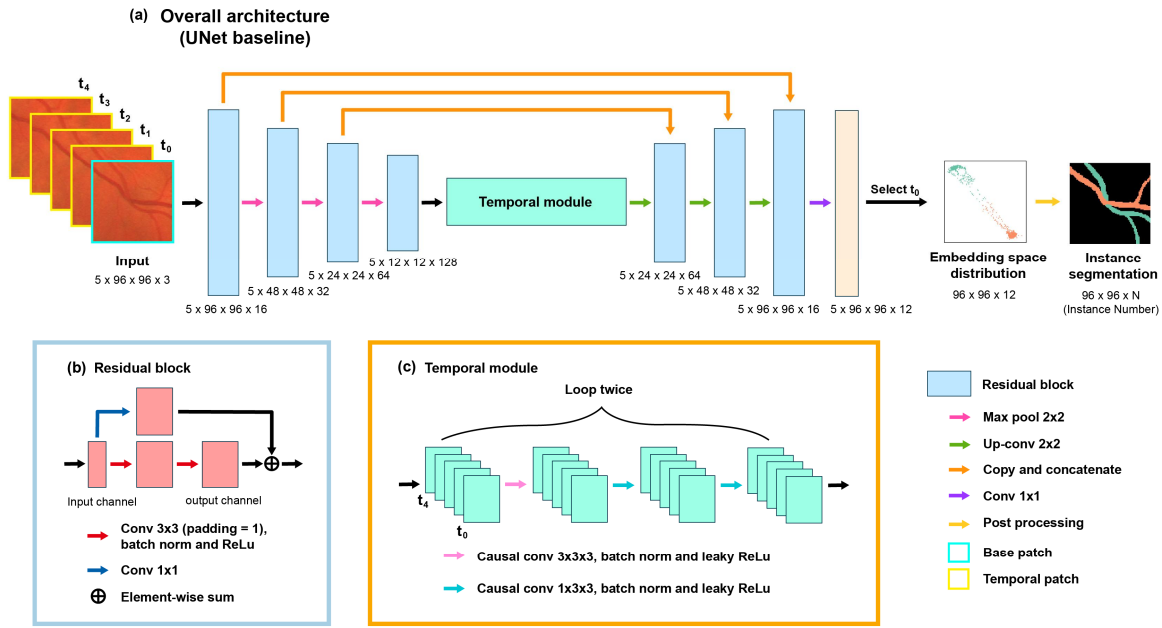

**Figure 2** Architecture of (a) InSegNN with the structure details of (b) residual block and (c) temporal module. Input patch images are stacked in the order from $t_4$ (earliest) to $t_0$ (base), which are cropped along the skeleton of vessels. The last frame of output is retained as the target for clustering, which is visualized as 2D scatter plot using Multidimensional scaling (MDS) method for dimension reduction. Instance segmentation labels are obtained by clustering embedding space distribution using mean-shift algorithm.

For temporal learning, we added a temporal module (**Fig. 2c**) to capture the vessels' spatial feature beyond the size of each patch, such that the model possesses "memory" as it traces each vessel tree. A sequence of overlapping image patches ($t_4$-$t_0$) tracing the vessel, like a video camera following the target, will provide a "temporal dimension" to the original 2D dimensions, which improves the performance of model.

The temporal module contains a series of 3D causal convolution layers, inserted into the bottleneck of InSegNN. This module learns both spatial and inter-frame information at the same time. Due to the causal convolution [22] (**Methods section 4.1.2 / Supplemental Fig. 1**), the latest patch $t_0$ are influenced by all the earlier patches ($t_4$-$t_0$) along the tracing.

After the parallel processing of encoder, 2D image patches with consecutive vessel contents (from $t_4$ to $t_0$) are stacked as the inputs of temporal module (**Fig. 2c**). The outputs will be split into 2D forms and put into decoder parallelly.

## 2.3 Instance segmentation on individual patches

Public fundus image dataset DRIVE [23] was used to test the performance of our embedding-based instance segmentation method. The set of 40 images in DRIVE dataset are equally divided into half and half for the training dataset and test dataset. Instance segmentation masks were manually annotated and validated by a clinical expert. For training and test, patches with pixel size of 96 x 96 were cropped from the images to enrich the dataset and avoid complicated topology commonly seen in a full frame fundus photo.

To generate the consecutive temporal patches for temporal module, skeletons of vessels were manually extracted to serve as the tracing paths (**Fig. 3a**). Image patches were cropped along the skeletons with step of 10 pixels length. **Figure 3b** shows the example frames from one sequence in fundus image (**Fig. 3a**). Corresponding ground truth maps for vessel instance segmentation are displayed in **Figure 3c**.

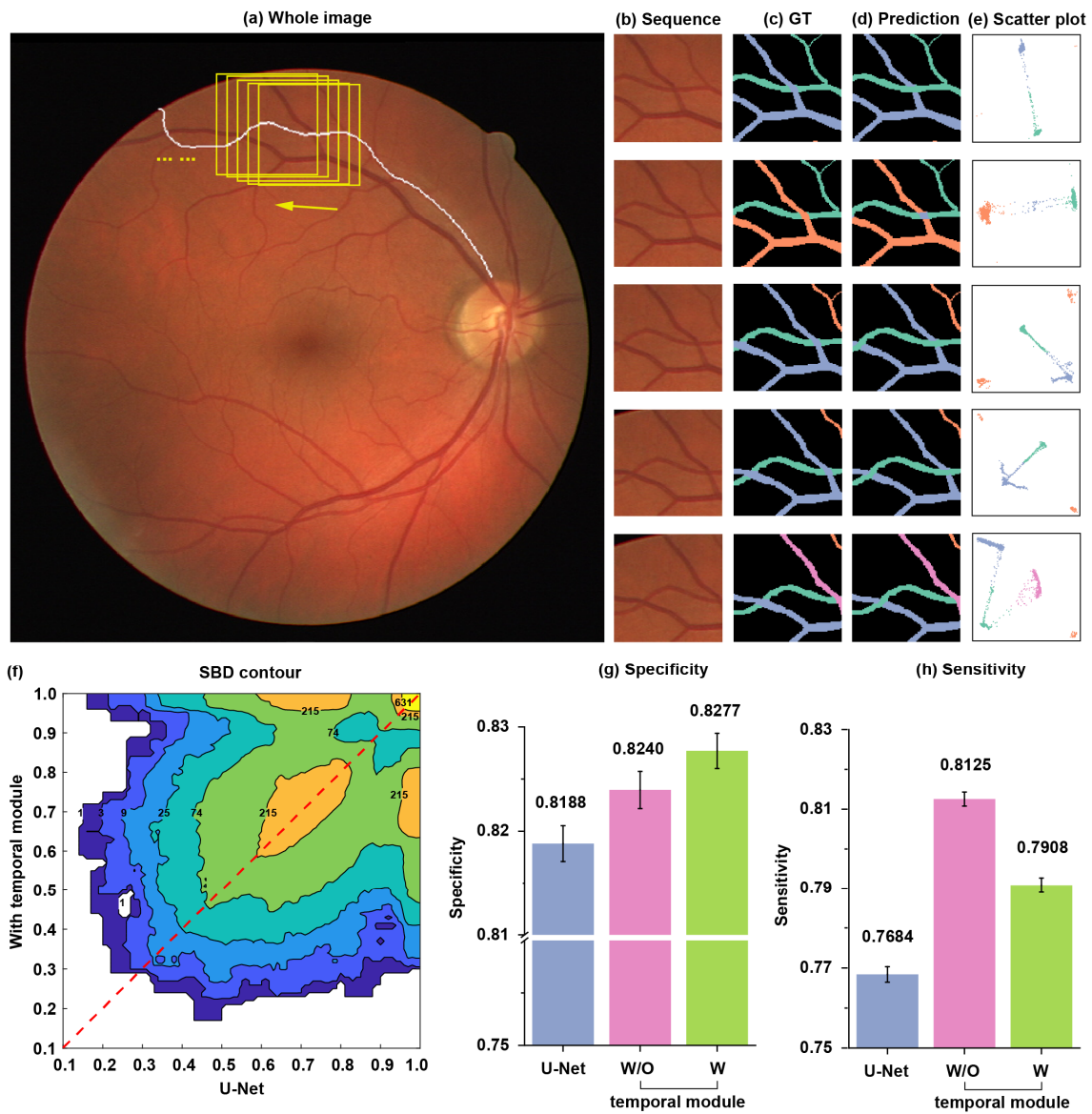

**Figure 3** Results of instance segmentation in the local region with aid of temporal learning. (a) Original fundus image. White line is the main skeleton of the vessel tree and temporal sequences are obtained by moving center of cropping window (yellow squares) through the skeleton. (b) Inputs. Patches are cropped at the locations of yellow squares from (a), stacked with 4 other neighbor patches and sent to the model. (c) Ground truth (GT). Different colors represent different instances. (d) Prediction by InSegNN with temporal module. Label colors are adjusted to correspond to ground truth. (e) Scatter plot of pixels in 2D embedding space. Dimension of original embedding space is lowered to 2D by MDS method for visual clarity. Colors of scatter points correspond to the labels of prediction (d). (f) Contour map recording SBD distribution of test datasets. X axis shows SBD of U-Net model while y axis is SBD of InSegNN with temporal module. Numbers on contour show the thresholds between levels. As the input of contouring, heatmap is generated with 100 bins, points on which are filtered by 8x8 mean filter with symmetric padding. (g) Bar plot of specificity for U-Net, InSegNN without and with temporal module. Error bars are standard errors of mean. (h) Bar plot of sensitivity for U-Net, InSegNN without and with temporal module. Error bars are standard errors of mean.

**Figure 3d** shows the prediction of the testing dataset by our InSegNN and confirms the vessel instances are accurately labeled. The corresponding scatter plot of all vessel pixels in the embedding space is shown in **Fig. 3e**. Pixels of different vessel instances are successfully clustered. Only a few pixels fall between clusters that mostly belong to overlaps from crossing vessel trees. Additional examples are provided in **Supplemental Fig. 2.**

To quantitively evaluate the performance, we used **Symmetric Best Dice** (SBD) specifically designed for instance segmentation [24, 25]. It measures the comprehensive similarity of prediction and ground truth among all instances. We used a standard U-net [26] with the same discriminative loss function as a baseline. **Figure 3f** plots SBD contour of our network against U-net on 7828 testing patches, where contour stretches more in upper-left region, indicating that SBD is improved over U-net. We further performed an ablation study to test the benefit of the temporal module in our network, by simply only using $t_0$ for training and testing. **Figure 3g** and **3h** shows the comparison between U-net, InSegNN without and with temporal module, on the sensitivity and specificity (**see definition in Method 4.2**) of the instance segmentation. Adding temporal module improves the specificity, i.e., identify vessel instances accurately, and reduces the sensitivity comparing without the temporal module.

## 2.5 Global vessel tracing performance

To complete vessel tracing over the whole image, we developed an algorithm constructed by two iterations (**Fig 4a**). The first iteration implemented aforementioned spatial multi-sampling, that a local spatial shift of the base patch $t_0$ (note: $t_1$-$t_4$ remained the same), inspired by "microsaccade", is applied within the inner iteration. The five instance segmentation labels were averaged to produce a probability map at base patch $t_0$. The outer iteration exhausts all the new starting points within each patch, until no new starting points found. Every iteration will also dynamically update the probability map, and thus further improve the robustness (**Supplemental Movie**).

**Figure 4b-4d** illustrated the performance of **spatial multi-sampling, and dynamic probability map (Fig. 4b-4d)**. As demonstrated in the previous section, temporal learning improves the accuracy of vessel instance segmentation (**Fig. 3f, 3g**). Still, the temporal learning alone is not sufficient, particularly at the far end of the vessel branching when the vessel contrast declines, or vessel trees extensively overlap (**Fig. 4b**). Instead of relying solely on instance segmentation per patch, we dynamically update the probability map for vessel tracing, by overlapping and averaging the instance segmentation as the patch traces along the vessel tree (**Supplemental Movie**). Because the patch moves to the new starting point found at the boundary of each patch, there is at least 25% overlapping areas and thus improves the robustness by building up the probability map. **Figure 4c** illustrated the probability map with dynamic iteration. Using a threshold of 0.6 on the map, the vessel tree was correctly identified and segmented.

When we further included spatial multi-sampling, by shifting the patch and repeated the instance segmentation 4 times (i.e. spatial multi-sampling), each pixel was sampled >=5 times. **Figure 4d** displays total 5 instance segmentations by InSegNN on original patch and 4 slight displacements of the patches by 10 pixels in upper, lower, left, and right directions. Thanks to spatial multi-sampling, the probabilities of incorrect pixels are reduced obviously in **figure 4d**, making binarization perform better.

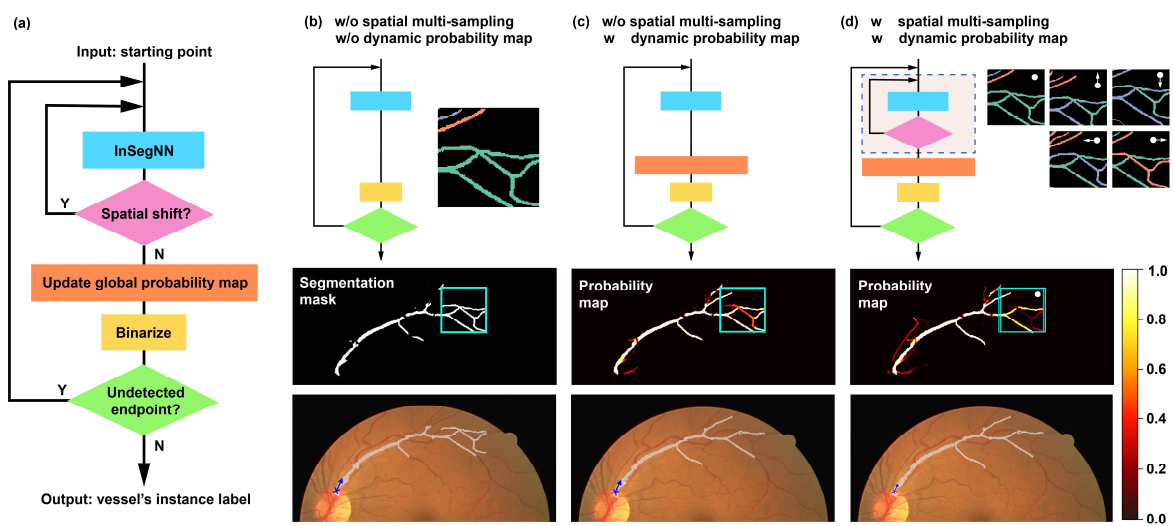

**Figure 4.** (a)Flow chart of vessel tracing. (b-d) Verifications of strategies for robustness in vessel tracing

process. Thumbnail flow charts in first row show difference with (a). (b) Control group without spatial multi-sampling and dynamic probability map . First row shows output of InSegNN, serving as part of segmentation mask in second row and final segmentation is shown in third row. (c) Group without spatial multi-sampling but with dynamic probability map. Local instance segmentation is same as (b) and it serves as a new sample updated in global probability map in second row. Third row is vascular mask segmented from probability map with a threshold of 0.6. (d) Group with spatial multi-sampling and dynamic probability map (standard vessel tracing process). First row shows 5 local instance segmentations achieved by spatial multi-sampling. White arrows mark the shifting direction. Base patch is labeled by white circle. Probability map in second row overlapping all 5 samples and are averaged and segmented as vascular mask in third row with the same threshold in (c). The color bar of probability map for (c) and (d) is shown on the right side of the whole row.

**Figure 5** shows two examples of global vessel tracing over the whole fundus images. The ground truth labels different vessel trees, and so does the tracing output from our methods. While some tracing ended early, presumably due to lack of vessel contrast in the original fundus images, different trees are distinguished accurately. More importantly, the hierarchy of the vessel branching is preserved by the tracing process (**Supplemental Movie**), as shown by gradient of the color for each tree.

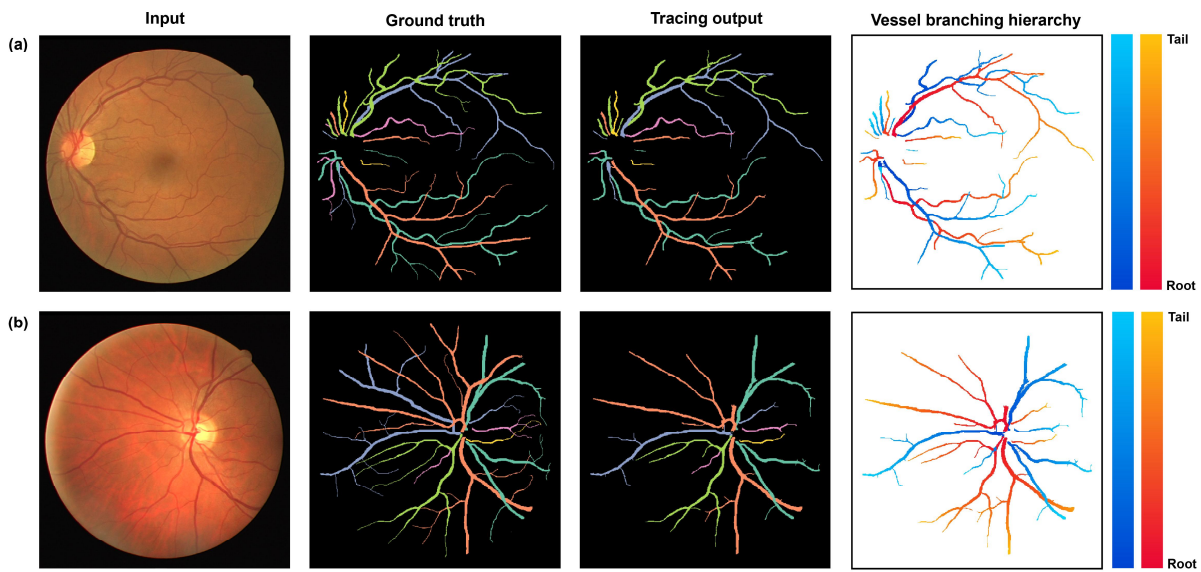

**Figure 5** Two test examples of the vessel tracing on fundus images. The first column is the original image. The second column is the ground truth, where labels with different colors or untouched with others represent different instances. The third column is the tracing output, whose label colors correspond to the ground truth. The fourth column is the vessel branching hierarchy where gradient colors annotate the branching directions. The color bars are shown on the right side. Two kinds of color bars are applied for visual clarity without any classification meaning.

# Discussion

This work presented a semi-automated vessel tracing method by vessel instance segmentation in a fundus retinal image. The essence of the method is a novel embedding-based model guided by the discriminative loss function in a neural network. This model, with aid of temporal learning, overperformed the baseline U-Net to produce local instance segmentation of vessels. To further increase the robustness of algorithm, spatial multi-sampling and dynamic iteration on global probability map are developed. Importantly, the nature of the tracing methods preserved the hierarchy of the vessel tree, laying down the solid foundation for the subsequent morphological quantifications.

For some typical errors or controversial cases (**Fig 6**) in instance segmentation, we can observe the pixel distribution in embedding space and find the inspirations for future improvement. It is worth noting that the current embedding-based method does not split pixel into two instances. Pixels of overlapping vessel regions are either randomly clustered into one of the crossing vessels, or sometimes assigned to a new instance (**Fig. 6a, 6e**). Isolated instance also becomes challenging to be segmented when the vessel contrast is low **(Fig. 6b)**. **Figure 6c** shows the inherent disadvantage of embedding space method when there is only one instance presented. Mean-shift algorithm mistakenly splits it into several instances once pixels scatter in embedding space unevenly and loosely. As for this phenomenon, we observed that our model with temporal module could improve the accuracy (**Supplemental Fig. 3-4**). **Figure 6d** reveals that small and vague vessels attached to large main vessels tend to be mislabeled as a new instance. It implies that the embedding-based neural network potentially learns to partition a large main instance into several sub-instances, creating a hierarchy structure, which is also observed by the work of multi-layered maps of neuropil [27]. Because this ability appears naturally, the segmentation of sub-instances is unstable and the distance between each sub-cluster is uneven, thus misleading the mean-shift algorithm to cluster some of them as new instances. InSegNN underperforms at very complicated topological situations. **Figure 6e** shows an example that one main vessel and two branches form a triangle and one bifurcation point almost overlaps the intersecting part of two main vessels. It is difficult to distinguish even for human experts, especially when some of the vessel segments lack contrast.

We envision two potential approaches can be explored to improve the instance segmentation quality. Firstly, unique features and topology of vessels could be paid attention as the entry point. Some research shows that the bifurcation angle of the vessels has a theoretical optimal value of 75 degrees in normal subjects [28]. Also, curvilinearity of vessels could be characterized and utilized by various shape algorithms [29], no matter in traditional methodologies or deep learning ways. Secondly, an alternative strategy is to consider the sub-instances (branches of vessels) as the fundamental units and recombine them by post processing or manual evaluation. Though it leads to the embedding-based method no longer an end-to-end one, this strategy can be effective when the vascular structure becomes further complicated and tortuous.

To summarize, for the first time to our knowledge, vessel tracing algorithm by instance segmentation is developed as a robust tool for retinal vasculature analysis in ophthalmology. Our work paves the way for down-stream vascular morphological and topographic analysis for a better understanding of pathophysiology in a broad range of retinal conditions.

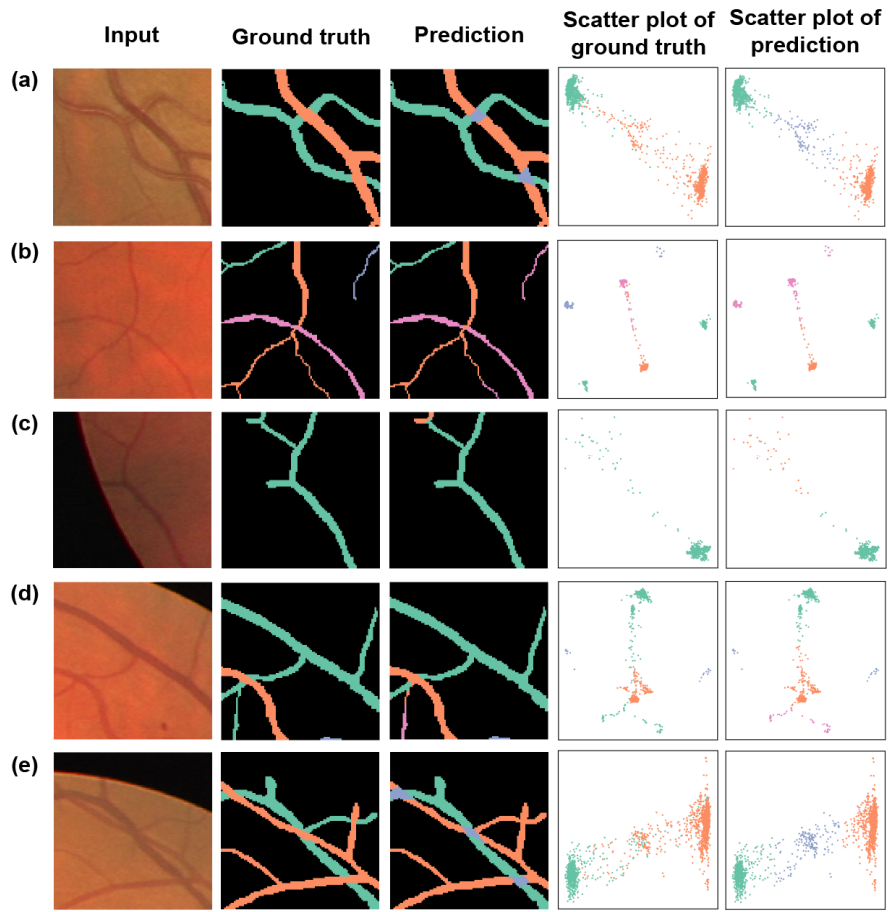

**Figure 6.** Examples of 5 typical errors shown in the embedding-based instance segmentation. (a) Overlap as a new instance; (b) isolated instance assimilated to other instances; (c) mis-segmentation in single instance case; (d) small branch as a new instance; (e) deficient performance of complex topological situations.

# METHODS

## 4.1 Model architecture

### 4.1.1 Discriminative loss function

Embedding-based network produces a representation space of the image, where each pixel has its own embedding vector. In the embedding space, pixels of the same instance cluster together and different clusters escape from each other. After sufficient training, the pixels in this representation space can be easily split into instances by some post-processing steps like *k*-means and mean-shift algorithm.

Discriminative loss function, specifically designed for instance segmentation using embedding method, describes three powerful forces that help separate clusters. The attraction force (**Eq. 1**) acts on each instance itself, attracting pixels of it approaching the center of clusters (mean embedding position of the pixel assembly), thus decreasing the variation. Meanwhile, the repulsion force (**Eq. 2**) is applied on every pair of clusters, pushing them away to keep a distance. The repulsion force enlarges the distance between clusters. The regularization force (**Eq. 3**) draws all clusters to the origin of the embedding space, ensuring the activation function is effective.

Assume that one image has $C$ instances, and each instance consists of $N_c$ pixels ($1 \leq c \leq C$). As a correspondence, in the embedding space, the cluster number is $C$ and the element number of each cluster is $N_c$. The discriminative loss function is defined as below:

$$L_{attraction} = \frac{1}{C}\sum_{c=1}^{C}\frac{1}{N_c}\sum_{i=1}^{N_c} max\left(0, \|\mu_c - x_i\| - \delta_v\right)^2, \tag{1}$$

$$L_{depulsion} = \frac{1}{C(C-1)}\sum_{c_A=1}^{C}\sum_{c_B=1, c_A \neq c_B}^{C} max\left(0, 2\delta_d - \|\mu_{c_A} - \mu_{c_B}\|\right)^2, \tag{2}$$

$$L_{regularization} = \frac{1}{C}\sum_{c=1}^{C}\|\mu_c\|, \tag{3}$$

$$L_{disc} = \alpha L_{attraction} + \beta L_{depulsion} + \gamma L_{regularization}, \tag{4}$$

where $\mu_c$ is the mean embedding position of each cluster and $x_i$ is the embedding position of pixel $i$. Variance threshold $\delta_v$ defines the maximum radius of one cluster and distance threshold $\delta_d$ is half of the minimum distance between two clusters. $\alpha$, $\beta$ and $\gamma$ control the weight of three forces, respectively. Ideally, if $\delta_v < \delta_d$ and $L_{disc} = 0$, clusters won't get in touch with others and thus, different instances can be segmented directly by clustering algorithms. In our experiments, we set hyper parameters $\delta_v = 0.5$, $\delta_d = 3$, $\alpha = 1$, $\beta = 1$ and $\gamma = 0.001$.

**4.1.2 Encoder-decoder structure**

Embedding-based instance segmentation, driven by discriminative loss function, is flexible to the selection of models. As shown in **Figure 2**, classical encoder-decoder structure is chosen as the base structure of InSegNN and temporal module is inserted into the bottleneck. U-Net's first layer dimension is set as 16 Inputs are first

The encoder and decoder undergo three rounds of down-sampling and up-sampling because the size of input is relatively small (96 x 96) and receptive fields are large enough after three times down sampling. In addition, the residual block (**Figure 2b**) replaces the basic backbone at each layer to allow information to flow from initial to last layers.

The encoder and decoder only make effect in spatial dimensions. The encoder adopts max-pooling layers to increase the receptive field and extract global features from vessel images. The channel of inputs, initially 3, firstly increases to 16 by residual module, and then doubles before every up-sampling. Then the decoder transforms features into embedding space vectors. The channel halves after every down-sampling. Skip connections between layers of the encoder and decoder are constructed to fully use different levels of features and replenish the lost information caused by max-pooling layers.

**4.1.3 Temporal module**

After the encoder, the temporal image sequence is stacked in a new dimension with length 5 and is put into the temporal module including a series of causal convolution layers followed by batch norm and leaky ReLu with negative slope of 0.01(**Figure 2c**). The causal convolution series are looped twice.

As shown in **Supplemental Fig. 1**, for a given time index $t$, the causal convolution layer only convolves frames with the indices equal to or smaller than $t$. Thus, the corresponding outputs aren't interfered by future frames, indicating it is causal and the last frame merges the information from all previous frames.

### 4.1.4 Output layer

After the decoder, a 2D convolution layer with 1x1 kernel is introduced to decrease the dimension of the output to 12. Semantic mask helps remove the non-vascular background before the final post-processing. Because a lot of state-of-art semantic vessel segmentation models [30-32] were proposed, and this work concentrated more on instance segmentation, we used the ground truth vessel mask provided by the dataset.

### 4.1.5 Adjusted temporal discriminative loss function

Due to the addition of the temporal module, one output consists of $T$ sets of embedding vectors, corresponding to $T$ patches of ground truth. The discriminative loss function is modified as below to fully take advantage of multi-temporal outputs:

$$L_{temporal} = \frac{1}{T}\sum_{t=1}^{T}(L_{disc})_t, \tag{5}$$

### 4.1.6 Post-processing

To ultimately achieve the instance segmentation, every pixel needs to be clustered into different instances according to the corresponding embedding space vectors provided by the neural network. Mean-shift algorithm was used for clustering and labeling. As an unsupervised learning algorithm, it has no requirement to know the instance number but tracks the cluster centers naturally by following the point distribution itself.

## 4.2 Metrics

**Specificity**, **sensitivity**, and **Symmetric Best Dice (SBD)** are evaluation metrics for the performance of neural network in instance segmentation task. In this work, they are constructed based on the original dice score (**Eq. 6**). SBD consists of two parts, one of which uses prediction as benchmark, i.e. specificity (**Eq. 7**), while other one considers from the perspective of ground truth, i.e. sensitivity (**Eq. 8**). For each instance in the prediction, SBD firstly computes the best dice score of it by traversing all instances from the truth to match it. Then an average best dice score of all instances in the prediction is calculated. The same process is applied to every instance in the truth symmetrically.

The final SBD is the minimum of the average best dice scores of the prediction and truth (**Eq. 9**). Specificity (prediction-side) concentrates on the accuracy of prediction, for example the fine details of edges. In contrary, sensitivity (truth-side) weights large-scale matching between ground truth and prediction instance.

$$Dice(Predict, Truth) = \frac{2TP + smooth}{2TP + FP + FN + smooth}, \qquad (6)$$

$$Specificity = \frac{1}{N_{predict}} \sum_{i=1}^{N_{predict}} \max_{1 \leq j \leq N_{truth}} \left( Dice(Predict_i, Truth_j) \right), \qquad (7)$$

$$Sensitivity = \frac{1}{N_{truth}} \sum_{j=1}^{N_{truth}} \max_{1 \leq i \leq N_{predict}} \left( Dice(Predict_i, Truth_j) \right), \qquad (8)$$

$$SBD = min(Specificity, Sensitivity), \qquad (9)$$

### 4.3 Vessel tracing algorithm

Vessel tracing algorithm, based on InSegNN, aims to segment all vessel instances on global retinal images. **Figure 4a** displays a general flow chart of algorithm while **Algorithm 1** reveals the details of algorithm. **Supplemental algorithm 2,** shows the generation of temporal sequence from a vessel tree while **Supplemental algorithm 3** specifically describes how to generate a vessel tree structure from instance segmentation mask.

For each vessel tree, the input starting point and direction vector are manually annotated by marking two points $P_1$, $P_2$ close to the source of the vessel tree. Make the $P_1$ close to the source and set it as the starting point and calculate the vector $P_2 - P_1$ to acquire the direction vector.

At the beginning of tracing, only one image is sent into the model due to the lack of temporal information, which is acceptable for InSegNN when predicting. As the iteration times increase, the temporal sequence length turns longer until reaching 5, whose detail can be viewed in **Supplemental algorithm 2**.

Instance segmentation mask of each vessel tree and corresponding vessel branching hierarchy are updated through iterations (**Supplemental algorithm 3**). Branching hierarchy, i.e. tree data structure, which records the node and connection relationship of vasculature, are utilized for detection of new starting points and shown as gradient of colors in accordance with the distances to the tree origin along the vessel centerline (**Fig 1b, Fig 5**).

**Algorithm 1** Vessel tracing

**Input:**
Image $I$, semantic segmentation mask $I_{semantic}$, starting points of each vessel trees $P_{start}$, direction vector $\vec{D}$, neural network **InSegNN**

**Output:**
Instance segmentation mask $I_{instance}$, tree data structure of vessels $T$

**For each vessel trees:**
1: **while** undetected $P_{end}$ found **do**
2:     **while** use spatial shifting **do**
3:         crop a 96 x 96 image patch whose center is the $P_{start} + shifting$
4:         **if** tree structure $T$ does not exist **then**
5:             only use this image patch as the input $S$
6:         **else**
7:             trace back to generate the temporal sequence $S$ (see **Supplemental algorithm 2**)
8:         **end if**
9:         input the $S$ to **InSegNN**, mask the results by $I_{semantic}$ and use mean-shift algorithm to output the instance labels
10:         add the label that contains $P_{start}$ to the global probability map
11:     **end while**
12:     binarize the probability map by threshold of 0.6 to acquire the current $I_{instance}$ and set $P_{start}$ as used one
13:     **if** tree structure $T$ does not exist **then**
14:         skeletonize the label and extract the end points $P_{end}$ of it (end points should be at the boundary of the image patch)
15:         remove the $P_{end}$ satisfying $P_{end} = \mathbf{argmin}\left(dot(P_{end} - P_{start}, \vec{D})\right)$, which is marked as the source of this vessel segment, namely $P_{source}$
16:         skeletonize the tree label and build the tree structure $T$ whose origin is $P_{source}$ (see **Supplemental algorithm 3**)
17:     **else**
18:         update the tree structure according to the vessel tree label (see **Supplemental algorithm 3**)
19:     **end if**
20:     Find $P_{end}$, i.e. branch tails of tree $T$, and set the $P_{end}$ closet to the old $P_{start}$ as new $P_{start}$
21: **end while**

Combine the instance labels and tree data structure from different vessel trees to output the final $I_{instance}$ and $T$

## 4.4 Experiment Setup

### 4.4.1 Dataset

DRIVE dataset contains 40 color fundus images including 7 abnormal pathology cases. Data were collected from a diabetic retinopathy screening program in the Netherlands. In our experiment, test dataset uses first half of images while training dataset uses second half. One of the training dataset images (34_training) was abandoned due to its low imaging quality. Images possess the size of 584 x 565 pixels with RGB channels. The dataset provides semantic segmentation masks, which are directly provided for training and test. Instance segmentation masks were manually annotated based on semantic segmentation mask and validated professionally. The instance segmentation masks remove the small vessel branches and segments that own unclear topological relationships with others. The instance segmentation masks do not distinguish the up-down relationship of overlaps. The intersection is added to both of masks.

A specific strategy of cropping image patches was adopted for temporal sequence. Firstly, the skeletons of vessels were extracted to serve as the tracing paths. Secondly, image patches were cropped following the skeleton with a 10 pixels step length. The center pixels of the image patches must be located at the center line of vessels, guaranteeing that at least one vessel would appear in a patch. During the training, 5 continuous patches are extracted from random parts of the sequence to constitute temporal sequence. The same step was processed on the test dataset. Statistics containing numbers of the subjects, sequences, patches, range and median of the Instance numbers per patch are listed in **Supplemental table 1.**

### 4.4.2 Training setting

The neural network was trained through 300 epochs with a learning rate of 0.0001 and a batch size of 64. During the training, Image patches were randomly flipped in horizontal and vertical direction, randomly rotated by 0, 90,180, 270 degrees and normalized with mean as 0 and standard deviation as 1 for data augmentation.

# Supplemental materials

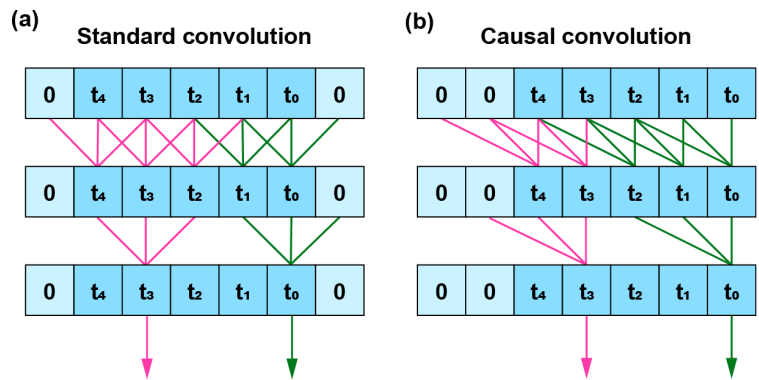

**Supplemental figure 1.** Comparison of (a) standard convolution and (b) causal convolution

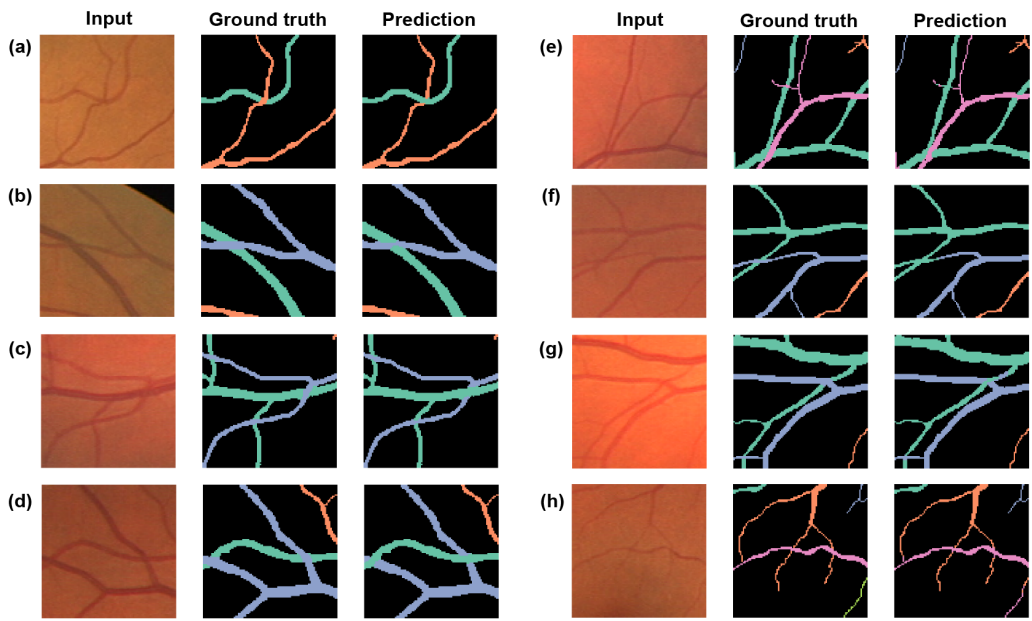

**Supplemental figure 2.** Results of the vessel instance segmentation in local region.

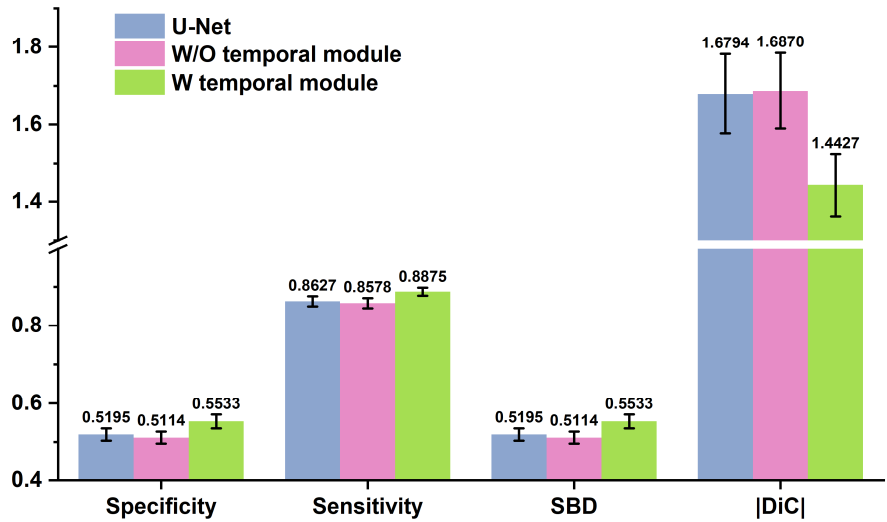

**Supplemental figure 3.** Bar plot of specificity, sensitivity, SBD and |DiC| (Absolute difference of count between prediction and truth) for U-Net, InSegNN without and with temporal module when only single instance cases (131 samples) in test datasets are gathered for statistics. Error bars are standard errors of mean. Our model with temporal module overperforms others evidently.

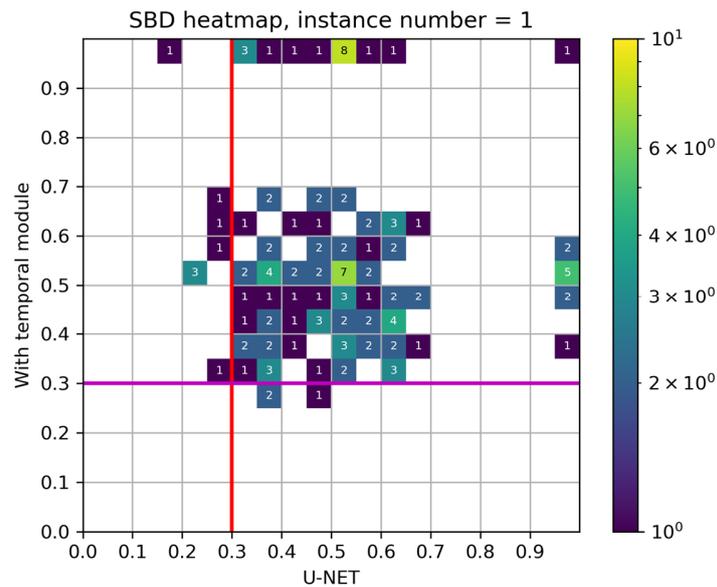

**Supplemental figure 4.** SBD heatmap of single instance cases (131 samples) in test datasets, shown in log scale. X axis shows SBD of U-Net model while y axis is SBD of InSegNN with temporal module. More cases increase score after changed to our model (left side of red line) compared to cases changed from our model to U-Net (region below magenta line), indicating our model has advantage dealing with single instance.

| Number | Training dataset | Test dataset |
|---|---|---|
| Subjects | 19 | 20 |
| Sequences | 507 | 579 |
| Patches | 7825 | 7828 |
| Sequences per subject *(min, max, median)* | 21, 37, 27 | 17, 40, 29.5 |
| Instances per patch *(min, max, median)* | 1, 13, 3 | 1, 18, 2 |

**Supplemental table 1.** Statistics of the training dataset and test dataset

**Supplemental movie** (shown by .mp3 in attachment)

**Supplemental algorithm 2** Acquire temporal sequence from a vessel tree

**Input:**
Tree data structure of vessels $T$, center point $P_{target}$

**Output:**
Temporal sequence $S$

1: search the node in $T$ to find $N_{target}$ closest to $P_{target}$
2: find $N_{target}$'s parent node $N_{parent}$ and their connected vessel centerline $C$
3: tracing from $N_{target}$ along $C$ with the step length 10 pixels to crop 4 patches
4: **if** $C$ is shorter than 40 pixels **then**
5:     find $N_{parent}$'s parent node and repeat the above step
6: **end if**
7: 4 patches are tacked with patch whose center is $P_{target}$ to constitute temporal sequence $S$

**Supplemental algorithm 3** Generate tree data structure

**Input:**
Instance segmentation mask $I_{instance}$, origin point of vessel tree $P_{origin}$

**Output:**
Tree data structure of vessels $T$, endpoints $P_{end}$ of all branches

1: skeleton map $I_{skeleton} = \mathbf{skeletonize}(I_{instance})$
2: node type map $I_{map} = (I_{skeleton} * \begin{bmatrix} 1 & 1 & 1 \\ 1 & 1 & 1 \\ 1 & 1 & 1 \end{bmatrix}) \odot I_{skeleton}$
3: set $P_{origin}$ as $P_{start}$
4: **for** each undetected neighbor point $P_{neighbor}$ of $P_{start}$:
5:     **if** $I_{map}[P_{neighbor}] = 2$ (endpoint of one branch) **then**
6:         stop, record the endpoint $P_{end}$ to tree data structure $T$ and return to last undetected $P_{neighbor}$
7:     **else if** $I_{map}[P_{neighbor}] = 3$ (connection point) **then**
8:         set this $P_{neighbor}$ as the new $P_{start}$ and return to **step 4**
9:     **else if** $I_{map}[P_{neighbor}] = 4, 5\ or\ 6$ (bifurcation point) **then**
10:         record the bifurcation point to tree data structure $T$ set this $P_{neighbor}$ as the new $P_{start}$ and return to **step 4**
11:     **end if**
12: Once all points are detected, end the algorithm, and output the tree data structure of vessels $T$, endpoints $P_{end}$ of all branches